\documentclass[doublecol]{epl2}
\usepackage{graphicx}

\title{Detectable inertial effects on Brownian transport through  narrow pores}

\author{Pulak Kumar Ghosh\inst{1,2} \and Peter H\"anggi\inst{2} \and Fabio Marchesoni\inst{3} \and Franco Nori\inst{1,4} \and Gerhard Schmid\inst{2}}
\shortauthor{P. K. Ghosh \etal}

\institute{
  \inst{1} Advanced Science Institute, RIKEN, Wako-shi, Saitama, 351-0198, Japan\\
  \inst{2} Institut f{\"u}r Physik, Universit{\"a}t Augsburg, D-86135 Augsburg, Germany\\
  \inst{3} Dipartimento di Fisica, Universit\`a di Camerino, I-62032 Camerino, Italy\\
  \inst{4}Physics Department, University of Michigan, Ann Arbor, MI
48109, USA
}

\abstract{
We investigate the transport of suspended Brownian particles dc
driven along corrugated narrow channels in a regime of finite
damping. We demonstrate that inertial corrections cannot be
neglected as long as the width of the channel bottlenecks is
smaller than an appropriate particle diffusion length, which
depends on both, the temperature and the strength of the dc drive.
Therefore, transport through sufficiently narrow constrictions
turns out to be sensitive to the viscosity of the suspension
fluid. Applications to colloidal systems are discussed.}

\begin{document}

\maketitle

Brownian transport in narrow corrugated channels is a
topic  of potential applications to both natural
\cite{Hille,Brenner} and artificial devices \cite{BM}. Depending
on the amplitude and geometry of the wall modulation, corrugated
channels fall within two distinct categories: (i) smoothly
corrugated channels, typically modeled as quasi one-dimensional
(1D) periodic channels with axial symmetry and unit cells
delimited by smooth, narrow bottlenecks, also called pores
\cite{chemphyschem,Zwanzig,Reguera:2001,Kalinay,Laachi,Reguera:2006,Burada,Reichelt,Jacobs};
(ii) compartmentalized channels \cite{Lboro,Bere1,PHfest,Shape},
formed by identical compartments separated by thin dividing walls
and connected by narrow pores centered around the channel axis.
Brownian transport in such sharply-corrugated channels must be
treated as an irreducible two or three dimensional diffusion
problem \cite{Schuss}. More importantly, for both categories of
corrugated channels most analytical results only apply under the
condition of very narrow pores \cite{Brenner,Zwanzig,Schuss}.

Corrugated channels are often used to  model transport of dilute
mixtures of small particles (e.g., biomolecules, colloids or
magnetic vortices) in confined geometries \cite{BM}. Each particle
is subjected to thermal fluctuations with temperature $T$ and
large viscous damping $\gamma$, and a homogeneous constant bias of
strength $F$ parallel to the channel axis.  Such a dc drive is
applied by coupling the particle to an external field (e.g., by
attaching a dielectric or magnetic dipole, or a magnetic flux to
the particle), without inducing drag effects on the suspension
fluid. Interparticle and hydrodynamic interactions are thus
ignored for simplicity (these assumption are discussed in Ref.
\cite{chemphyschem}).

\begin{figure}[tp]
\centering

\includegraphics[width=0.45\textwidth]{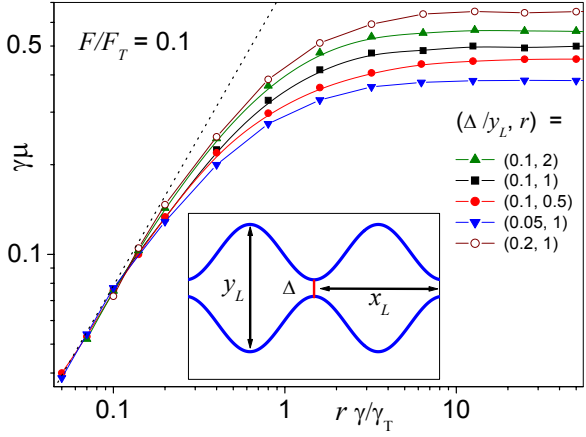}
\caption{(Color online) Rescaled mobility, $\gamma \mu$, versus rescaled damping constant,  $r \gamma/\gamma_T$,
in the corrugated channel of Eq.~(\ref{walls}) at small drive,
$F/F_T=0.1$, and for different pore sizes, $\Delta/y_L$, and
aspect ratio, $r=x_L/y_L$. All quantities are expressed in dimensionless units with scaling parameters $r=x_L/y_L$,
$F_T=kT/\Delta$ and $\gamma_T=\sqrt{mkT}/\Delta$, see Eq.~(\ref{gT}). The dotted line is the fitting law of Eq.~(\ref{th-mu0}). The
actual parameter values used in our simulations are $m=kT=y_L=1$.
Inset: corrugated channel with the boundary function $w(x)$ given in Eq.~(\ref{walls}).
\label{F1}}
\end{figure}

In this paper we investigate the relevance of inertial effects due
to the viscosity of the suspended particle. As is often the case
with biological (and most artificial) suspensions \cite{BM}, the
Brownian particle dynamics in the bulk can be regarded as
overdamped. This corresponds to (i) formally setting the mass of
the particle to zero, $m=0$, or, equivalently, to make the
friction strength $\gamma$ tend to infinity, and (ii) assuming $F$
smaller than the thermal force $F_0=\gamma \sqrt{kT/m}$
(Smoluchowski approximation) \cite{SEQ}. The current literature on
corrugated channels invariably assumes such an overdamped limit.
But how large is an ``infinite" $\gamma$ (or how small can be a
``zero" $m$)? The answer, of course, depends on the geometry of
the channel. Our main conclusion is that the overdamped dynamics
assumption for Brownian diffusion through pores of width $\Delta$
subjected to a homogeneous drive $F$, applies only for $\gamma \gg
\sqrt{mkT}/\Delta$, and $\gamma \gg \sqrt{mF/\Delta}$,
irrespective of the degree of corrugation. This means that the
inertial effects cannot be neglected as long as the Brownian
diffusion is spatially correlated on a length
($l_T=\sqrt{mkT}/\gamma$ at small dc drive, or $l_F=mF/\gamma^2$
at large  dc drive) of the order of, or larger than, the pore
width $\Delta$. Therefore, for sufficiently narrow pores or
sufficiently large drives, inertia always comes into play by
enhancing the blocking action of the channel bottlenecks. Thus,
the condition of vanishingly narrow pores, $\Delta \to 0$, assumed
in most analytical studies, can be inconsistent with the
assumption of overdamped diffusion in the spirit of the
Smoluchowski approximation \cite{SEQ}. Finally, we also discuss
applications to transport in colloidal systems.

Let us consider a point-like Brownian particle diffusing  in a two
dimensional (2D) suspension fluid contained in a periodic channel
with unit cell $x_L \times y_L$, as illustrated in Fig.~\ref{F1} (inset). The particle is subjected to a homogeneous force ${\vec F} = F {\vec e_x}$ oriented along the $x$-axis. 
The damped Brownian particle obeys the 2D Langevin equation,
\begin{equation}\label{le}
m{d^2{\vec r}}/{dt^2}=-\gamma ~{d{\vec r}}/{dt}+{\vec F}\;+
\sqrt{2\gamma kT}~{\vec \xi}(t),
\end{equation}
where ${\vec r}=(x,y)$. The random forces ${\vec
\xi}(t)=(\xi_x(t),\xi_y(t))$ are zero-mean, white Gaussian noises
with autocorrelation functions $\langle \xi_i(t)\xi_j(t')\rangle =
\delta_{ij}\delta(t-t')$, with $i,j=x,y$. The symmetric walls of
the corrugated channel have been modeled by the sinusoidal
function $\pm w(x)$ (Fig.~\ref{F1}, inset),
\begin{equation}
w(x)=(1/4)[(y_L+\Delta)-(y_L-\Delta)\cos(2\pi x/x_L)].
\label{walls}
\end{equation}
With this choice for $w(x)$ we made explicit contact with the
current literature on entropic channels
\cite{chemphyschem,Zwanzig,Reguera:2001,Kalinay,Laachi,Reguera:2006,Burada}.
However, similar results were obtained for sharper corrugation
profiles, e.g., for compartmentalized channels
\cite{Lboro,Bere1,PHfest,Shape} (not reported here). We
numerically integrated Eq.~(\ref{le}) by a Milstein algorithm
\cite{Milstein} with time-step, $\Delta t$, ranging from $10^{-6}$
down to $10^{-9}$, as $F$ was increased. For each run, $\Delta t$
was set small enough for the output to be independent of it. The
stochastic averages reported below were obtained as ensemble
averages over 10$^6$ trajectories with random initial conditions;
transient effects were estimated and subtracted.

\begin{figure}[tp]
\centering
\includegraphics[width=0.45\textwidth]{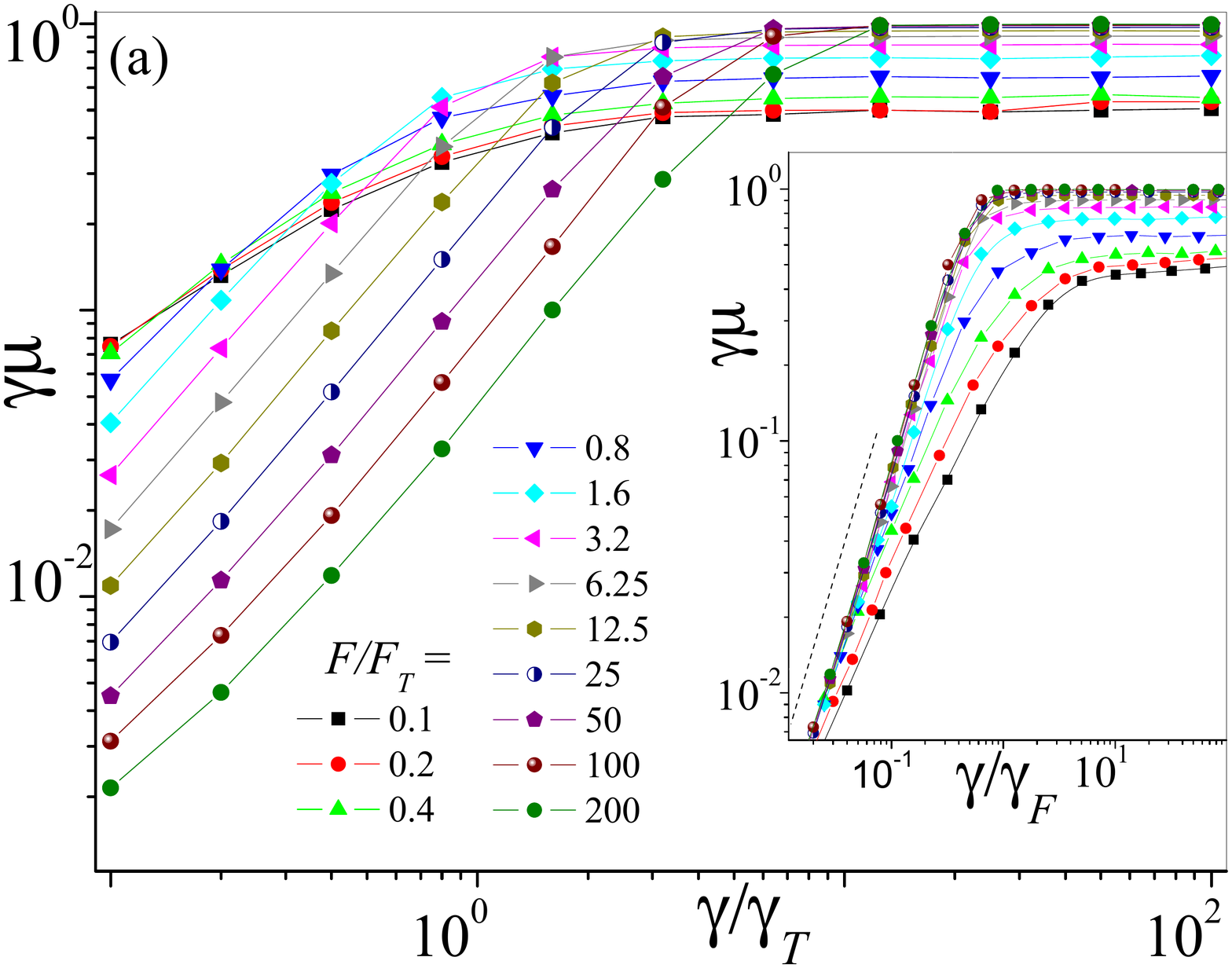}
\includegraphics[width=0.45\textwidth]{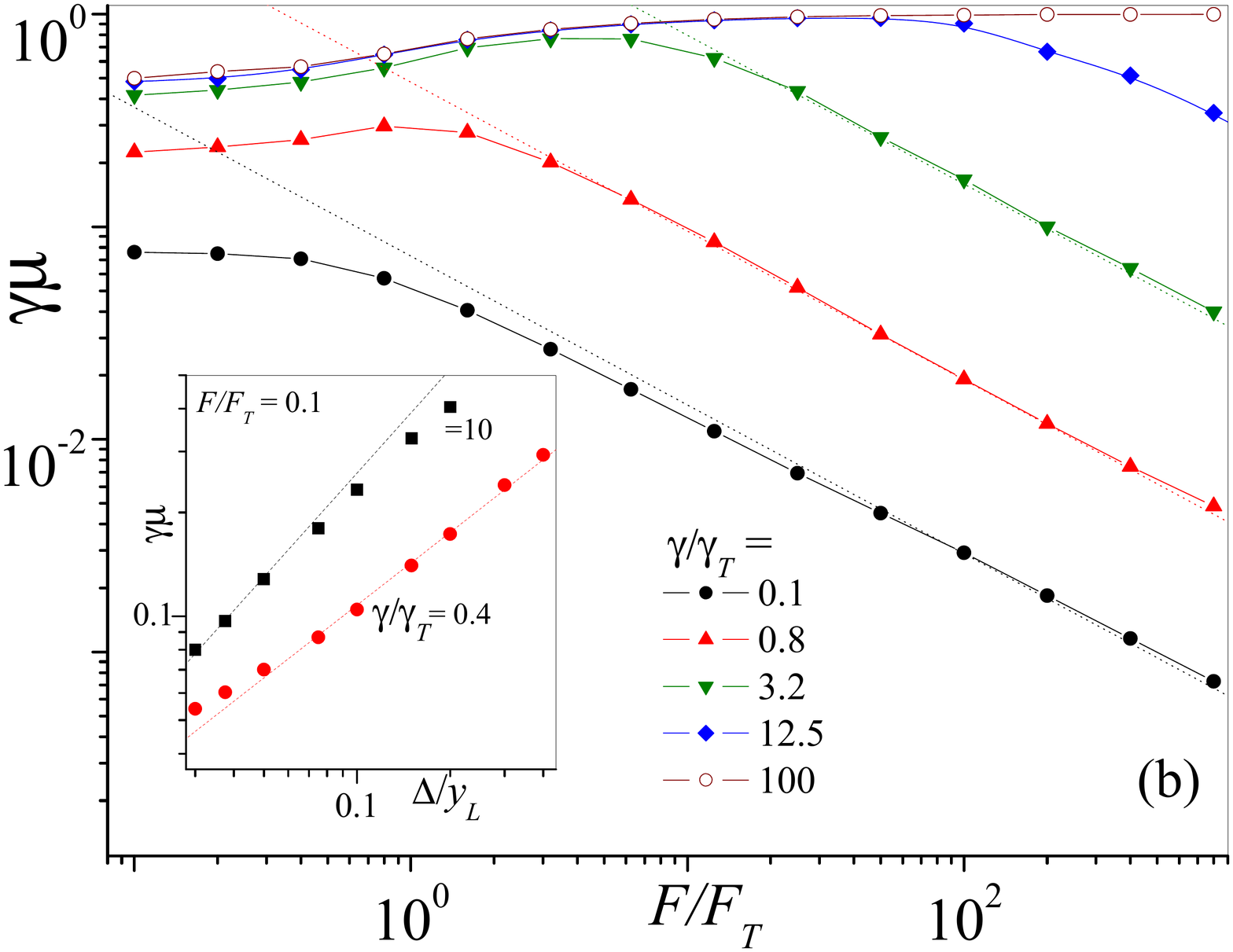}
\caption{(Color online) Rescaled mobility, $\gamma \mu$, in a
corrugated channel with aspect ratio $r=1$, rescaled pore size $\Delta/y_L=0.1$, and (a) {\it vs.}
the rescaled damping $\gamma/\gamma_T$ for different force  values $F$; (b) {\it vs.}
$F/F_T$ for various $\gamma$.  The scaling parameters are $F_T=kT/\Delta$ and $\gamma_T=\sqrt{mkT}/\Delta$;  the actual
simulation parameters were set to $m=kT=y_L=1$, as in Fig. \ref{F1}. Inset (a): data from the main panel after different
rescaling of damping constant, $\gamma/\gamma_F$, with $\gamma_F=\sqrt{mF/\Delta}$, see Eq. (\ref{gF}); the raising branch of the universal curve for large
drives is fitted by the power law $(\gamma/\gamma_F)^\alpha$ with
$\alpha=1.4$. Inset (b): $\gamma \mu$ {\it vs.} $\Delta/y_L$ for
$\gamma/\gamma_T=0.4$ and two values of $F$. The dashed lines in
(b) represent the large-drive power laws $(F/F_T)^{-\alpha/2}$
(main panel) and $(\Delta/y_L)^{\alpha/2}$ (inset), both with
$\alpha=1.4$; for $F/F_T=1$, $\mu$ is proportional to $\Delta$
(inset). \label{F2}}
\end{figure}

Two quantifiers have been used to characterize the Brownian
transport in such sinusoidally corrugated channel:

{\it (i) mobility.} The response of a Brownian particle in a
channel  subjected to a dc drive, $F$, oriented along the
$x$-axis, is expressed by the mobility, $ \mu(F) = \langle v(F)
\rangle /F, $ where $\langle v\rangle \equiv \langle \dot x(F)
\rangle =\lim_{t\to\infty}[\langle x (t) \rangle -x(0)]/t$. The
function $\mu(F)$ increases from a relatively small value for
$F=0$, i.e. $\mu(F=0) = \mu_0$, up to the free-particle limit,
$\gamma \mu_\infty=1$, for $F\to \infty$ \cite{Burada}. We recall
that in the bulk, a free particle drifts with speed $F/\gamma$.

{\it (ii) diffusivity.} As a Brownian particle is driven across a
periodic array of bottlenecks or compartment pores, its
diffusivity, $ D(F) = \lim_{t \to \infty}[\langle x^2(t)\rangle -
\langle x(t) \rangle^2]/2t, $
picks up a  distinct $F$ dependence. In corrugated channels with
smooth bottlenecks, for $F\to \infty$ the function $D(F)$
approaches the free or bulk diffusion limit, $D(\infty)=D_0\equiv
kT/\gamma$, after going through an excess diffusion peak centered
around an intermediate (temperature dependent \cite{Burada}) value
of the drive.  Such a peak signals the depinning of the particle
from the barrier array of $w(x)$ associated with the channel
bottlenecks \cite{Costantini}.

Recall that,  in the absence of external drives and for any value
of the damping constant, Einstein's relation, $ \gamma \mu_0 =
D(F=0)/D_0, $ establishes   the dependence of the transport
parameters on the temperature and the channel compartment geometry
under equilibrium conditions. 

Inertial effects  in corrugated channels become apparent both for
small $\gamma$ and for large $F$. By inspecting Figs. \ref{F1} and
\ref{F2} we immediately realize that (for small $\gamma$) inertia
tends to suppress the particle mobility through the channel
bottlenecks. At large $F$, when plotted versus $\gamma$ [Fig.
\ref{F2}(a)], the rescaled mobility approaches unity for $\gamma
\to \infty$, as expected in the Smoluchowski approximation
\cite{Burada}, but drops to zero in the underdamped limit, $\gamma
\to 0$. More remarkably, the resulting $\gamma \mu$ curves shift
to higher $\gamma$ on increasing $F$ (main panel). On expressing
$\gamma$ in units of $\gamma_F=\sqrt{mF/\Delta}$, see Eq.
(\ref{gF}) below, the curves for large drives tend to collapse on
a universal curve well fitted by the power law
$(\gamma/\gamma_F)^\alpha $ with $\alpha=1.4$ (inset).
Correspondingly, in Fig. \ref{F2}(b) the mobility grows like
$\Delta^{\alpha/2}$ at large $F$ (inset) and decays like
$F^{-\alpha/{2}}$ for small $\Delta$ (main panel).

The power law, $\gamma \mu \propto (\gamma/\gamma_F)^\alpha$,
introduced here is only a convenient fit of the rescaled mobility
function, even if it holds for two or more decades of
$\gamma/\gamma_F$. The analytical form of that function remains to
be determined. Deviations from the fitted power law, $\mu \propto
\Delta^{\alpha/2}$, for the mobility at very small $\Delta$ [Fig.
\ref{F2}(b), inset], suggest that the fitting exponent, $\alpha$,
slightly depends on $\Delta$, with $\alpha \to 2$ in the limit
$\Delta \to 0$ (not shown). [Note that, on the contrary, the power
law $\gamma \mu \propto F^{-\alpha/2}$ works throughout the entire
$F$ range explored in Fig. \ref{F2}(b).]

The dependence of the rescaled mobility on the system parameters
in the underdamped limit is further illustrated in Fig. \ref{F1},
where at low $\gamma$ and for vanishingly small drives, the
mobility grows proportional to the aspect ratio $r=x_L/y_L$ of the
channel unit cell and the pore cross section $\Delta$, i.e.,
scales with the dimensionless quantity $r\gamma/\gamma_T$ [see
also the data set for $F/F_T=1$ in the inset of Fig. \ref{F2}(b)].

\begin{figure}[tp]
\centering
\includegraphics[width=0.45\textwidth]{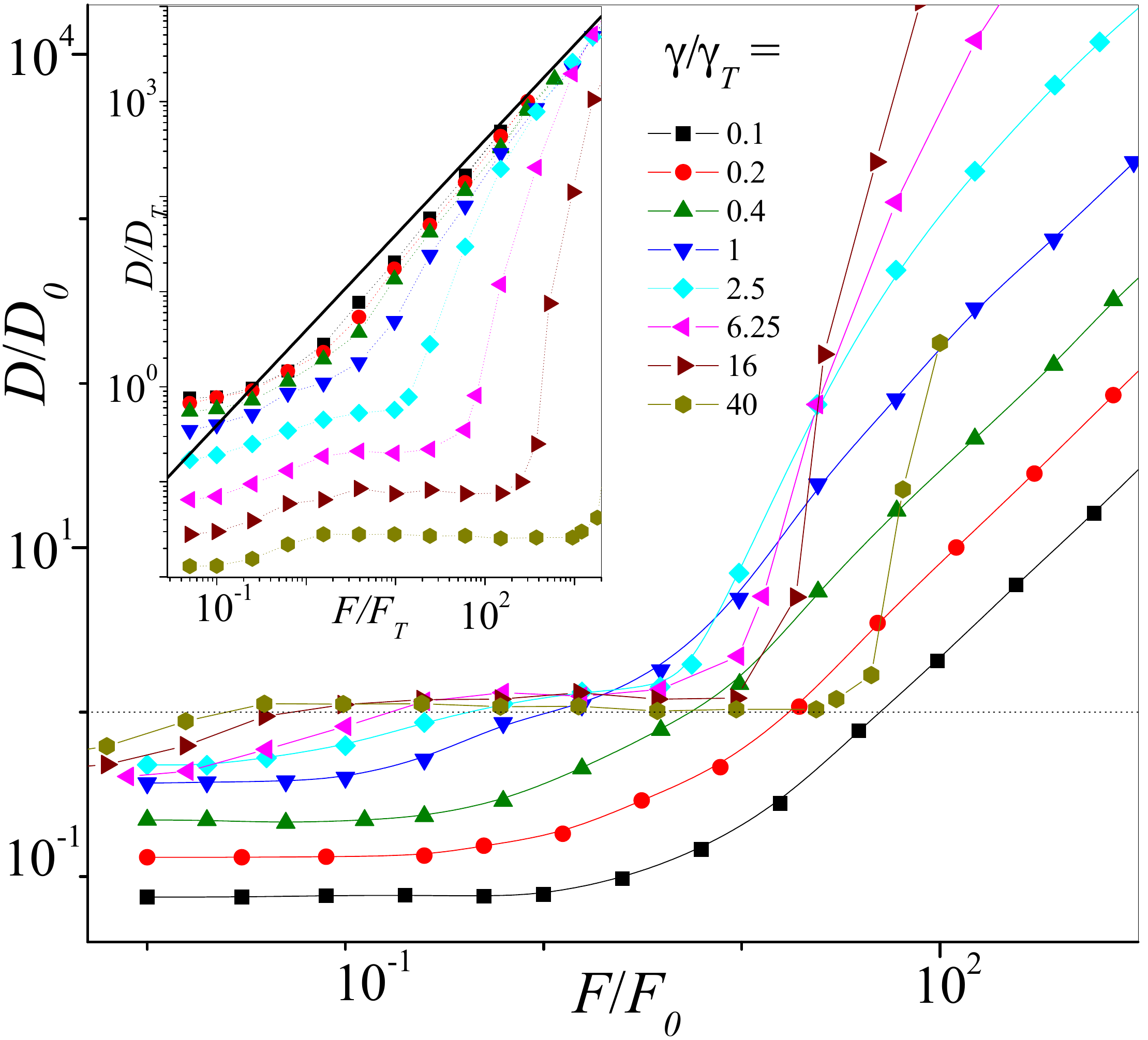}
\caption{(Color online) Rescaled diffusivity, $D/D_0$, {\it vs.}
rescaled force, $F/F_0$, (main panel) and the rescaled diffusion coefficient $D/D_T$ {\it vs.} $F/F_T$ (inset) in the
corrugated channel of Eq. (\ref{walls}) with aspect ratio $r=1$, rescaled pore width
$\Delta/y_L=0.1$, and different friction coefficients $\gamma$. The actual simulation
parameters are $m=kT=y_L=1$. The meaning of the scaling parameters, $D_T= kT/\gamma_T$, $F_0=\gamma \sqrt{kT/m}$, and $F_T = kT/\Delta$, is discussed in the text. The solid line in the inset is the
heuristic power law of Eq. (\ref{corrugated-diff}). \label{F3}}
\end{figure}

Deviations of  the diffusivity data from the expected overdamped
behavior are even more prominent. As shown in Fig. \ref{F3}, at
large $\gamma$ the curves $D(F)$ approach the horizontal asymptote
$D(F)=D_0$, as expected \cite{Burada}. However, beyond a certain
value of $F$, seemingly proportional to $\gamma^2$ (see figure
inset), these curves abruptly depart from their horizontal
asymptote. In the underdamped limit, the $F$ dependence of the
diffusivity bears no resemblance with the typical overdamped
behavior. At low $\gamma$, all $D(F)$ data sets collapse to a
unique curve [Fig.~\ref{F3}, inset], which tends to a value
smaller than $D_0$ for $F\to 0$, and diverges for $F \to \infty$,
like $F^\beta$ with $\beta\simeq 1$. Such power law holds for
large $\gamma$, as well, though in the large $F$ domain, only.
Indeed, for exceedingly large $F$, all $D(F)$ curves seem to
eventually approach a unique asymptote, irrespective of $\gamma$.

By comparing the plots of Figs.~\ref{F1}-\ref{F3} we conjecture
that corrections due to inertia become significant in two regimes:\\
{(i)} at {\it low drives} for
\begin{equation}
\gamma ~^{\;<} _{\;\sim} ~\gamma_T=\sqrt{mkT}/\Delta. \label{gT}
\end{equation}
This characteristic damping was used to rescale the mobility data
in Fig. \ref{F1} [see also Fig. \ref{F2}(b), inset]; moreover, in
Fig. \ref{F3}, for $\gamma <\gamma_T$ the diffusivity becomes a
monotonic function of $F$ with no plateau around $D_0$. The
physical meaning of $\gamma_T$ is simple. For $\gamma<\gamma_T$
the {\em thermal} length $l_T=\sqrt{mkT}/\gamma$ grows larger than
the width of the pores, $\Delta$, so that the Brownian particle
cannot reach the normal diffusion regime, implicit in Einstein's
relation, before bouncing off the pore walls. As a consequence,
the Smoluchowski approximation fails in the vicinity of the
bottlenecks.

Replacing $\gamma$ with $\gamma_T$ in the bulk quantities $D_0$
and $F_0$ yields, respectively, $D_T=kT/\gamma_T$ and
$F_T=kT/\Delta$. These are the $\gamma$-independent rescaling
factors introduced in Figs. \ref{F1}-\ref{F3} to characterize the
inertia effects of the pore constrictions;
\\
{(ii)} at {\it high drives}, for
\begin{equation}
\gamma~^{\;<} _{\;\sim} ~\gamma_F=\sqrt{mF/\Delta}\;. \label{gF}
\end{equation}
In the presence of a strong dc drive, the condition $\gamma \gg
\gamma_T$ does  not suffices to ensure normal diffusion; the
additional condition  $\Delta>l_F$ is required. Here,
$l_F=mF/\gamma^2$ represents the {\em ballistic} length of a
driven-damped particle,
which is an estimate of the bouncing amplitude of a driven particle against the bottleneck. 
Upon increasing $F$ at constant $\gamma$, $l_F$ eventually grows
larger than $\Delta$ and inertia comes into play. This mechanism
is clearly responsible for the abrupt increasing branches of
$D(F)$ in Fig. \ref{F3}.

An analytical derivation of the transport quantifiers in the
presence of strong inertial effects (low $\gamma$ and/or large
$F$) proved a difficult task. However,  the following
phenomenological arguments  can explain well our numerical
findings.

Let us first consider the rescaled mobility at low drives. For $F=0$ the transport quantifiers $\gamma \mu_0$ and $D(0)$ can be formally expressed in terms of the mean exit time, $\bar \tau_e$, of the Brownian particle out of a single compartment, namely, $D(0)=x_L^2/4{\bar \tau_e}$ and $\mu_0=D(0)/kT$ (Einstein's relation). An analytical expression for $\bar \tau_e$ as a function of the compartment geometry is only available in the Smoluchowski approximation \cite{Schuss}. 
In the low damping regime, we only have a rough estimate of $\bar
\tau_e$. For $F=0$ the particle bounces off a compartment wall
with rate $(2/\pi)(2\sqrt{kT/m}/x_L)$ (attack frequency), but only
a fraction $\Delta/y_L$ of such collisions leads to pore crossing.
As a consequence, ${\bar \tau_e} \sim
4x_Ly_L\sqrt{m/kT}/\pi\Delta$ and
\begin{equation}
\gamma\mu_0\sim\frac{\pi}{4}\frac{\gamma
x_L}{\sqrt{kT}}\frac{\Delta}{y_L} = \frac{\pi}{4}\left
(r\frac{\gamma}{\gamma_T}\right ), \label{th-mu0}
\end{equation}
which closely reproduces the numerical data of Fig.~\ref{F1} for
$F/F_T=1$ with mobility $\mu \propto \Delta$.

The divergence of the diffusivity with the drive shown in
Fig.~\ref{F3}, $D(F) \propto F^\beta$ with $\beta=1$, is
surprising since in the Smoluchowski approximation $D(F)$ always
tends to its bulk value, $D_0$. A heuristic explanation for this
asymptotic power law runs as follows. The dispersion of a strongly
driven particle at low damping is due to its bouncing back and
forth inside the channel cells with speed $\pm \bar v$ and $\bar
v=F/\gamma$. Therefore, $D(F)={\bar v}^2{\bar \tau}_e/4$, where
${\bar \tau}_e$ is the particle mean exit time through one of the
pores of a channel cell, ${\bar \tau}_e=(\pi/2)(x_L/\bar v)
(\gamma/\gamma_T)$. The time constant ${\bar\tau}_e$ is estimated
here as the cell crossing time, $x_L/{\bar v}$, multiplied by the
geometric factor $\pi/2$ [due to the sinusoidal profile $w(x)$,
also used to derive Eq. (\ref{th-mu0})] and the success
probability, $\gamma/\gamma_T$, for the particle to actually cross
the bottleneck during a sequence of correlated bounces extending
over the relaxation time $m/\gamma$. One thus concludes that
$D(F)=(\pi/8)(x_L\Delta F)/\sqrt{mkT}$, or, in rescaled units of
$\gamma_T$ and $D_T$,
\begin{equation}\label{corrugated-diff}
\frac{D(F)}{D_T}=\frac{\pi}{8}\frac{x_L F}{kT},
\end{equation}
in rather good agreement with our simulation data (see inset of
Fig. \ref{F3}). Lowering the temperature, for small damping $D(F)$
diverges like $T^{-1/2}$, which means that diffusion is dominated
by a (chaotic) mechanism of ballistic collisions.

The main result of this work is that for real physical suspensions
flowing through compartmentalized geometries, both in biological
and artificial systems, the limit of vanishingly small pore size
becomes extremely sensitive to the finite viscosity of the
suspension fluid. With respect to previous attempts at incorporating finite-mass effects in the analysis of Brownian transport through corrugated narrow channels \cite{inertia},
we stress that the inertial effects reported here are not of mere academic interest. Indeed, such effects can become appreciable even at low Reynolds numbers and for relatively small drives, where the viscous action of the fluid on a suspended particle is well described by the Stokes'term in Eq. (\ref{le}) -- i.e., hydrodynamic corrections are negligible.

Inertial effects can be directly observed, for instance, in a
dilute solution of colloidal particles driven across a porous
membrane or an artificial sieve \cite{Mark}. To be specific, let
us consider spherical polystyrene beads \cite{Garbow} of radius
$r_0=1\mu$m, suspended in a low viscous medium with, say,
$\eta = 8.9 \times 10^{-4}$ Pa$\cdot$s (water), $6.0 \times
10^{-4}$ Pa$\cdot$s (benzene) or $3.6 \times  10^{-4}$ Pa$\cdot$s
(acetone). 
For the typical mass density of polystyrene, 200 kg/m$^3$, the
mass of a spherical bead of radius $r_0=$1 $\mu$m is $m \simeq 8.4 \times \; 10^{-16}\;$kg.
The corresponding rescaled damping constant $\gamma/m$, Eq.~(\ref{le}), can be determined by making use of Stokes' law, $\gamma/m = 6\pi \eta r_0/m$,
namely $\gamma/m \simeq 2.0 \times 10^{-5}$ ps$^{-1}$ (water),
$1.4 \times 10^{-5}$ ps$^{-1}$ (benzene) and $8.1 \times 10^{-6}$
ps$^{-1}$ (acetone).

At room temperature, $T=300$ K, the thermal energy unit is $kT =
4.14 \times \;10^{-21}\;$ kg$\,$m$^2$/s$^{2}$. If the pore radius
is, say, 10$\%$ larger than $r_0$, then the effective width
$\tilde\Delta$ for the finite-radius particle to go through the
pore, becomes $\tilde \Delta=\Delta-2r_0=0.2\;\mu$m and the
rescaled zero-drive damping threshold is
$\gamma_T/m=\sqrt{kT/m\tilde\Delta^2}$ is $1.1 \times 10^{-8}$
ps$^{-1}$. Here, $\gamma_T \ll \gamma$ justifying the applicability of
the overdamped dynamics. Nevertheless, for effective widths in the order of nanometers, i.e.
$\tilde\Delta = 0.002\;\mu$m, $\gamma_T/m \simeq 2.2 \times 10^{-6}$
ps$^{-1}$ reaches the order of the rescaled damping constant of acetone reported above.

Moreover, the corresponding non-scaled large-drive damping threshold
$\gamma_F/m=\sqrt{F/m\tilde\Delta}$ can be easily achieved when applying dielectrophoretic forces. Depending on
the size of the electrodes, dielectrophoretic forces acting on particles with radius 1$\; \mu$m can result in pulling forces of up to  $100$ pN \cite{Morgan}. As a consequence, $\gamma_F/m \simeq 1.1 \times 10^{-5}$ ps$^{-1} > \gamma_F/m$.
For suspended particles of even larger radius or higher
mass density, both conditions $\gamma<\gamma_T$ and $\gamma<\gamma_F$ are also
experimentally achievable.

In conclusion, the experimental  demonstration of inertial effects
on Brownian transport through narrow pores is easily accessible by
manipulating artificial particles of moderate size by means of
well established experimental techniques (see also \cite{extraref}). For too small particles,
like biological molecules, detecting such effects might require
more refined experimental setups.

\acknowledgments
This work was partly supported  by the European Commission under
grant No. 256959 (NANOPOWER) (FM), the Volkswagen foundation
project I/83902 (PH, GS), the German excellence
cluster''Nanosystems Initiative Munich'' (NIM) (PH, GS),  the
Augsburg center for Innovative Technology (ACIT) of the University
of Augsburg (FM, PH), and the Japanese Society for Promotion of
Science (JSPS) through Fellowship No. P11502 (PKG) and No. S11031
(FM). FN is partially supported by the ARO, NSF grant No. 0726909,
JSPS-RFBR contract No. 12-02-92100, Grant-in-Aid for Scientific Research (S), MEXT Kakenhi on Quantum Cybernetics, and the JSPS via its FIRST program.

\end{document}